\documentclass[aps,prl,reprint,superscriptaddress,floatfix]{revtex4-1}

\usepackage[pdftex]{graphicx}
\usepackage{color}
\usepackage{enumitem}
\usepackage{gensymb}
\usepackage{epstopdf}
\usepackage{bm}% bold math
\usepackage{amsmath,amsthm,amssymb}
\usepackage{dcolumn}% Align table columns on decimal point

\begin{document}

\title{Evolution of helimagnetic correlations in Mn$_{1-x}$Fe$_x$Si with doping: a small-angle neutron scattering study}
\author{L.J. Bannenberg}
\affiliation{Faculty of Applied Sciences, Delft University of Technology, Mekelweg 15, 2629 JB Delft, The Netherlands}
\email{l.j.bannenberg@tudelft.nl}
\author{R.M. Dalgliesh}
\affiliation{ISIS neutron source, Rutherford Appleton Laboratory, STFC, OX11 0QX Didcot, United Kingdom}
\author{T. Wolf}
\affiliation{Institute for Solid State Physics, Karlsruhe Institute of Technology, 76131 Karlsruhe, Germany}
\author{F. Weber}
\affiliation{Institute for Solid State Physics, Karlsruhe Institute of Technology, 76131 Karlsruhe, Germany}
\author{C. Pappas}
\affiliation{Faculty of Applied Sciences, Delft University of Technology, Mekelweg 15, 2629 JB Delft, The Netherlands}

\date{\today}

\begin{abstract}
We present a comprehensive small angle neutron scattering study of the doping dependence of the helimagnetic correlations in Mn$_{1-x}$Fe$_{x}$Si. The long-range helimagnetic order in Mn$_{1-x}$Fe$_x$Si is suppressed with increasing Fe content and disappears for $x$ $>$ $x^*$ $\approx$ 0.11, i.e. well before $x_C$ $\approx$ 0.17 where the transition temperature vanishes. For $x$ $>$ $x^*$, only finite isotropic helimagnetic correlations persist which bear similarities with the magnetic correlations found in the precursor phase of MnSi. Magnetic fields gradually suppress and partly align these short-ranged helimagnetic correlations along their direction through a complex magnetization process.
\end{abstract}

\maketitle

%############################################################################
%############################################################################
\section{Introduction}
%############################################################################
Tuning the interactions in magnetic materials by pressure or chemical substitution is a well-visited route to discover new and exotic phases of condensed matter. In chiral magnetism, the most notable example is provided by the effect of hydrostatic pressure on the properties of the archetype chiral magnet MnSi. In this system, the helimagnetic order at ambient pressure results from the competition between the ferromagnetic exchange and the Dzyaloshinsky-Moriya (DM) \cite{D,M} interaction that arises from the non-centrosymmetric crystal structure of this B20 compound \cite{bak1980}. The helical order, of which the propagation vector is fixed to the $\langle 111 \rangle$ crystallographic directions by magnetic anisotropy, is suppressed under pressure at $p_C$ $\approx$ 1.4~GPa. Above $p_C$, partial magnetic order persists in a non-Fermi liquid phase that emerges without quantum criticality \cite{pfleiderer2004,pfleiderer2007,uemura2007}. Additionally, topological contributions to the Hall effect hint that magnetic correlations with non-trivial topology, similar to the skyrmion lattice phase at ambient pressure \cite{muhlbauer2009,neubauer2009}, are stabilized in this region of the phase diagram \cite{ritz2013}.

Another way of tuning the chiral magnetic order is by chemically substituting MnSi with iron. In Mn$_{1-x}$Fe$_x$Si, the helimagnetic order is suppressed with increasing Fe concentration, with the spontaneous magnetization \cite{bauer2010,bannenberg2018mnfesisquid}, Curie-Weiss \cite{bannenberg2018mnfesisquid} and  transition temperature $T_C$ \cite{bauer2010,bannenberg2018mnfesisquid} extrapolating to 0 at $x_C$ $\approx$ 0.17. Remarkably, a change of magnetic behavior is already observed at $x^*$ $\approx$ 0.11 by magnetization \cite{bannenberg2018mnfesisquid}, magnetic susceptibility \cite{bannenberg2018mnfesisquid}, resistivity \cite{demishev2013}, and electron spin resonance (ESR) measurements \cite{demishev2014}. Based on these results, it has been suggested that $x^*$ is a candidate for a quantum critical point (QCP), possibly associated with the suppression of the long-range helimagnetic order \cite{demishev2013,glushkov2015,demishev2016a,demishev2016b}. However, despite all these studies, the nature of $x^*$ and of the magnetic correlations for $x^*$ $<$ $x$ $<$ $x_C$ remain unclear.

In the following we address these points and discuss the helimagnetic order in Mn$_{1-x}$Fe$_{x}$Si around $x^*$. With small angle neutron scattering (SANS), we systematically study the evolution of the helimagnetic correlations as a function of both temperature and magnetic field. In particular, we investigate the effect of dilution and compare the topology of the magnetic correlations for $x$ = 0, 0.03, 0.09, 0.10, i.e. for $x$ $<$ $x^*$, with that for $x$ = 0.11, 0.14, i.e. for $x$ $>$ $x^*$. All measurements were performed by systematically applying the magnetic field both perpendicular and parallel to the incoming neutron beam. In this way, we obtain an overview of the topology of the helimagnetic correlations, both perpendicular and parallel to the magnetic field, in a way that is not provided by previous studies. 

The results show that with increasing Fe concentration, the helices at zero magnetic field first reorient from $\langle 111 \rangle$ to the $\langle 110 \rangle$ crystallographic directions and that the long-range helimagnetic order in Mn$_{1-x}$Fe$_x$Si disappears at $x^*$. For $x$ $>$ $x^*$ finite isotropic helimagnetic correlations set-in and which bear similarities to those seen in the precursor phase in MnSi \cite{grigoriev2005,pappas2009,janoschek2013,pappas2017}. Magnetic fields gradually suppress and only partly align the helices along their direction through a complex magnetization process.

%############################################################################
\section{Experimental}
%############################################################################
Single crystals of Mn$_{1-x}$Fe$_{x}$Si with nominal Fe concentration $x$ = 0.03, 0.09, 0.10, 0.11, and 0.14 were grown using the Bridgeman method. The composition of the single crystals, which are listed in Table \ref{table}, was checked with a PANalytical Axios x-ray fluorescence spectrometer and revealed Fe concentrations of 0.032, 0.089, 0.101, 0.112, and 0.140, respectively. The samples originate from exactly the same batches as the samples of our previous magnetization and susceptibility study \cite{bannenberg2018mnfesisquid}. They have irregular shapes and their dimensions vary from $\sim5\times5\times5$~mm$^3$ to $\sim10\times10\times15$~mm$^3$. The measurements on MnSi were performed on the same cubic single crystal with dimensions $\sim5\times5\times5$~mm$^3$ used in previous experiments \cite{pappas2017,bannenberg2017reorientations,sadykov2018}. The structure of all single crystals was checked with x-ray Laue diffraction and the $x$ = 0, 0.03, 0.10, 0.11 and 0.14 samples were aligned with the [$\bar{1}$10] crystallographic direction vertical. The $x$ = 0.09 sample was aligned with the [001] direction vertical.

The SANS measurements were performed on the time-of-flight instrument Larmor of the ISIS neutron spallation source using neutrons with wavelengths of 0.09 $\, \leq \lambda  \leq \,$1.25~nm. The samples were placed at a distance of 4.4~m from the detector that consists of 80 $^3$He tubes, each 8 mm wide. The SANS patterns were normalized to standard monitor counts and background corrected using a high temperature  measurement. The magnetic field was applied by a 3D vector cryomagnet either parallel ($\vec{H} || \vec{k}_i$) or perpendicular ($\vec{H} \perp \vec{k}_i$) to the incoming neutron beam designated by its wavevector $\vec{k}_i$. All measurements were performed by first zero field cooling the sample to the lowest temperature of interest. Then a magnetic field was applied and the signal was recorded by stepwise increasing the temperature. The system was brought to thermal equilibrium before the measurement at each temperature commenced. The fitted values reported in this paper have been obtained using the non-linear least squares method and the error bars correspond to one standard deviation.

\begin{table}[tb]
\centering
\caption{Overview of the Mn$_{1-x}$Fe$_{x}$Si compositions studied. The nominal composition, $x_{nom}$, was verified with x-ray fluorescence spectroscopy ($x_{XRF}$). The critical temperature $T_C$ has been determined from magnetic susceptibility \cite{bannenberg2018mnfesisquid} and for $x$ $\leq$ $x^*$ from the inflection point of the temperature dependence of the total scattered intensity at zero magnetic field. The pitch of the helical modulation $\ell$ is tabulated for $T$ = 2.5~K and $T$ = $T_C$. }
\label{table}
\vspace*{5mm}
\begin{tabular}{p{1.20cm}p{1.20cm}p{1.25cm}p{1.25cm}p{1.3cm}p{1.3cm}}
\hline \hline
$x_{nom}$    & $x_{XRF}$ & $T_C$ {[}K{]} (SQUID)  & $T_C$ {[}K{]} (SANS) & $\ell_{2.5\text{K}}$ [nm] & $\ell_{T_C}$ [nm] \\ \hline
0    & 0         & 29.2 & 28.8          & 18.2          & 16.0          \\
0.03 & 0.032     & 19.2 & 19.2          & 13.8          & 13.2          \\
0.09 & 0.089     & 8.1  & 7.8           & 9.5           & 9.7           \\
0.10 & 0.101     & 5.4  & 5.5           & 8.9           & 9.0           \\
0.11 & 0.112     & 5.0  & -             & 8.4           & 8.5     \\     
0.14 & 0.140     & 2.4  & -             & 7.0           & 7.0     \\ \hline\hline  
\end{tabular}
\end{table}

%############################################################################
\section{Zero Magnetic Field}
%############################################################################

%############ Zero Field Patterns ###########################################
\begin{figure*}[tb]
\begin{center}
\includegraphics[width= 1\textwidth]{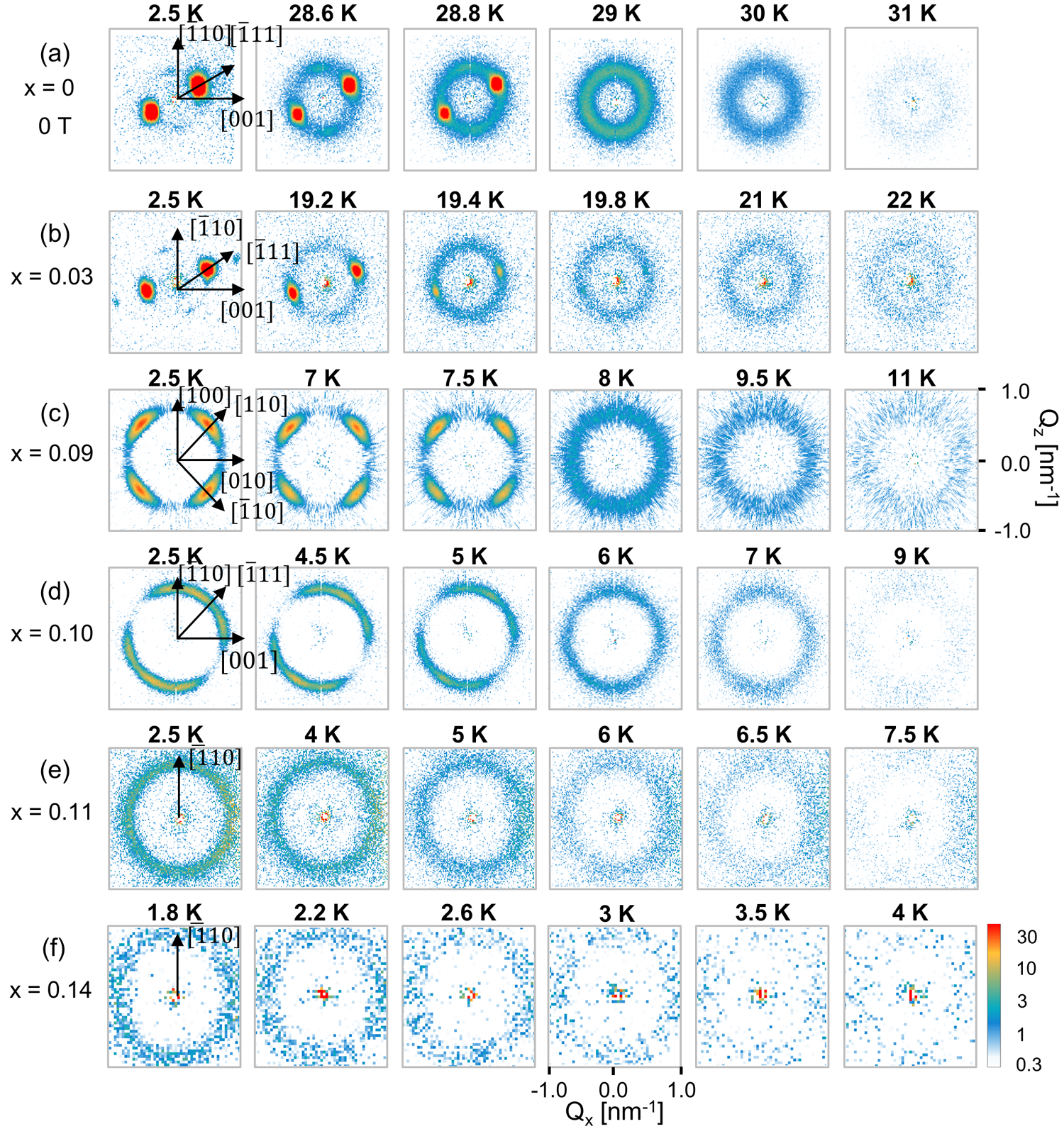}
\caption{SANS patterns recorded at zero magnetic field and different temperatures for Mn$_{1-x}$Fe$_x$Si with (a) $x$ = 0, (b) $x$ = 0.03, (c) $x$ = 0.09, (d) $x$ = 0.10, (e) $x$ = 0.11 and (f) $x$ = 0.14. }
\label{Zero_field_patterns}
\end{center}
\end{figure*}
%############################################################################

%%############  Zero Field Intensity #############################################
%\begin{figure}[ht!]
%\begin{center}
%\includegraphics[width= 0.4 \textwidth]{Intensity_Zero_field.png}
%\caption{Temperature dependence of the total scattered intensity obtained by summing the scattered
%intensity of the entire detector for the Mn$_{1-x}$Fe$_x$Si compositions indicated in the legend and at zero magnetic field. For clarity, the curves have been scaled by different arbitrary values.}
%\label{Zero_field_intensity}
%\end{center}
%\end{figure}
%%############################################################################

%############ Zero Field SQ ###########################################
\begin{figure}[ht!]
\begin{center}
\includegraphics[width= 0.35\textwidth]{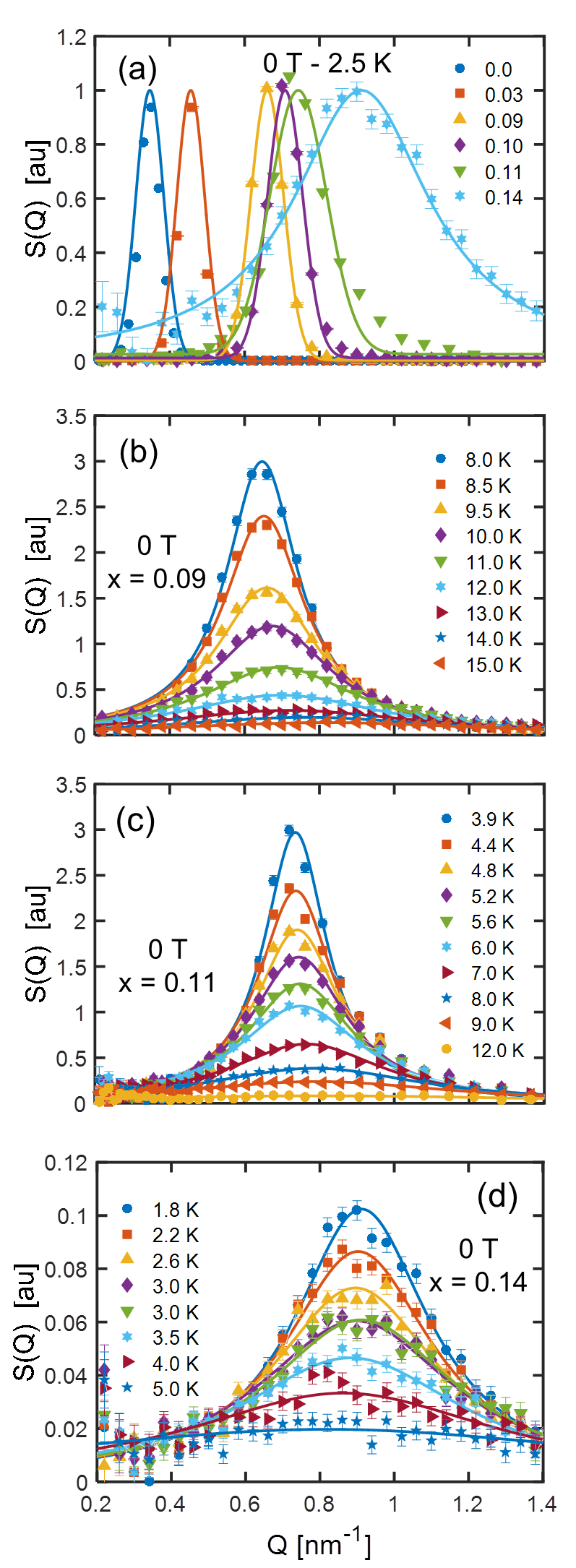}
\caption{The scattering function $S(Q)$, deduced by radially averaging the scattered intensity and at zero field for Mn$_{1-x}$Fe$_x$Si. In Panel (a) S(Q) is determined for $T$ $<$ $T_C$ and the Fe concentrations indicated in the legend. $S(Q)$ was measured at $T$ = 2.5~K for $x$ $\leq$ 0.11 and $T$ = 2.0~K for $x$ = 0.14 and is normalized to its maximum value. Panels (b) - (d) show $S(Q)$ at the indicated temperatures for (b) $x$ = 0.09, (c) $x$ = 0.11 and (d) $x$ = 0.14. In Panel (a) the solid lines indicate the best fits of the data to a Gaussian for $x$ $<$ 0.11 where the width of the Gaussian is fixed by the instrumental resolution. For $x$ $\geq$ 0.11 and in Panels (b)-(d) the solid lines represent the best fits of eq. \ref{eq:OZ} (convoluted with the instrumental resolution) to the data.}
\label{S_Q}
\end{center}
\end{figure}
%############################################################################

%############  Zero Field Fits #############################################
\begin{figure}[ht!]
\begin{center}
\includegraphics[width= 0.4 \textwidth]{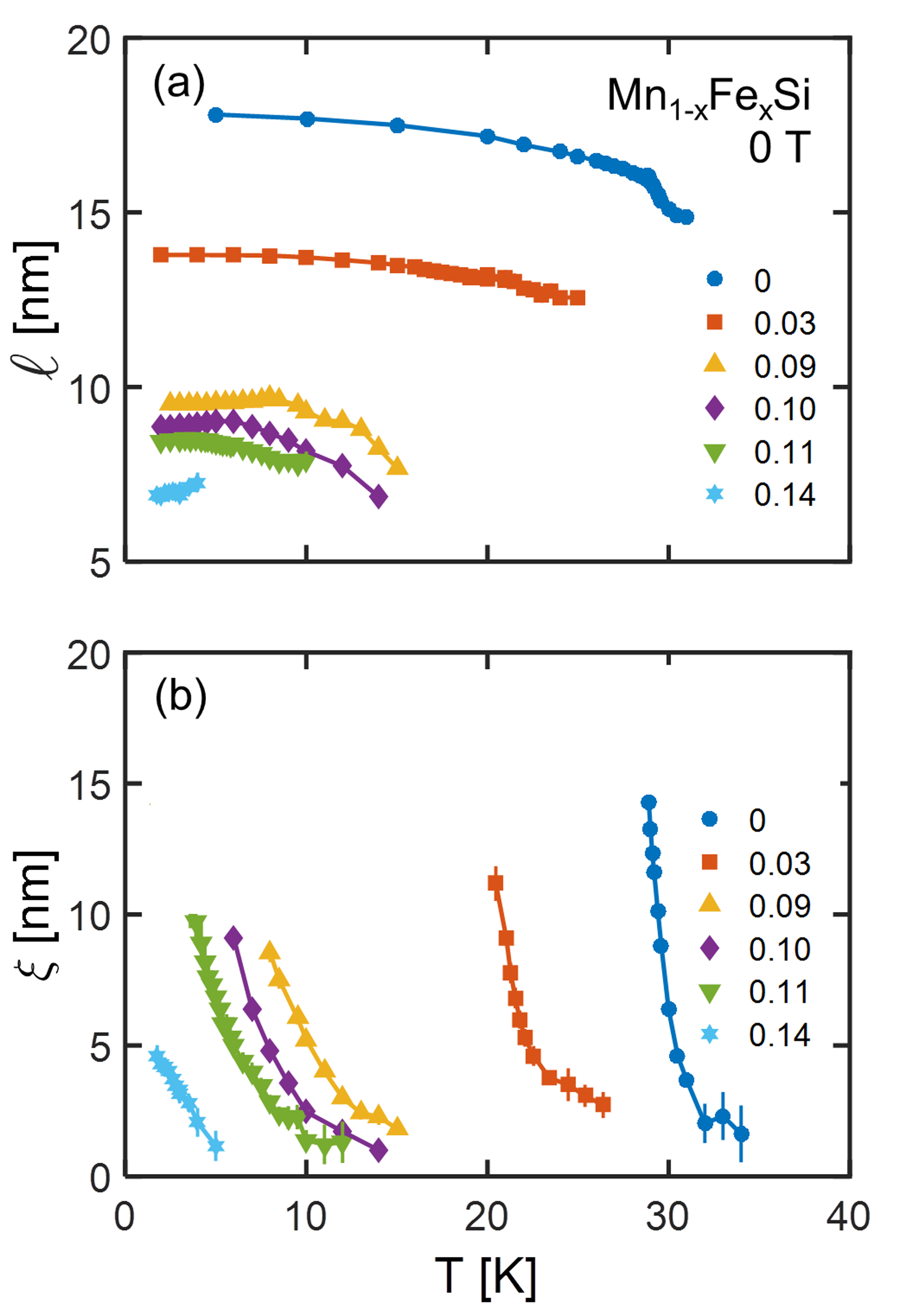}
\caption{Temperature dependence of (a) the pitch of the helical modulation $\ell$ and (b) the magnetic correlation length $\xi$  at zero magnetic field and for the Mn$_{1-x}$Fe$_x$Si compositions indicated in the legend. The values of $\xi$ are obtained by fitting $S(Q)$ to equation \ref{eq:OZ}. For $x$ $\leq$ 0.10 and $T$ $<$ $T_C$, $\ell$ was obtained from fitting $S(Q)$ to a Gaussian centered at $\tau=2\pi/\ell$. For $T$ $<$ $T_C$ and $x$ $\leq$ 0.10, the linewidth of $S(Q)$ is limited by the experimental resolution and $\xi$ cannot be determined.}
\label{Fit_SQ}
\end{center}
\end{figure}
%############################################################################

Figure \ref{Zero_field_patterns} depicts typical SANS patterns recorded at zero magnetic field and for different temperatures and compositions. The results for MnSi are displayed in Fig. \ref{Zero_field_patterns}(a) and are in good agreement with the literature \cite{grigoriev2005,grigoriev2007,janoschek2013,pappas2017}: above $T_C$, a diffuse isotropic ring of scattering  appears with radius $\tau=2\pi/\ell$, where $\ell$ is the pitch of the helical modulation. This ring, which originates from  isotropic, chiral and fluctuating helimagnetic correlations, intensifies and narrows when approaching $T_C$$\sim$ 28.7~K. This ring coexists in a temperature region of $\Delta T$ $\approx$ 0.2~K around $T_C$ with two Bragg peaks. These Bragg peaks mark the onset of the long-range helical order oriented along the $\langle 111 \rangle$ directions. The patterns displayed in Fig. \ref{Zero_field_patterns}(b) for $x$ = 0.03 show a qualitatively similar behavior, but with a lower transition temperature of $T_C$ $\sim$ 19.2~K. 

By further increasing the Fe concentration, the propagation direction of the helix below $T_C$ changes from  $\langle 111 \rangle$ to $\langle 110 \rangle$, as can be inferred from the alignment of the Bragg peaks along the $\langle 110 \rangle$ directions for $x$ = 0.09 [Fig. \ref{Zero_field_patterns} (c)] \footnote{The change of the propagation direction of the helix for $x$ = 0.09 is confirmed by triple axes spectroscopy measurements.}. This change of propagation direction is accompanied with a broadening of the Bragg peaks, which are no longer well-defined spots as for $x$ = 0 and 0.03 [Fig. \ref{Zero_field_patterns}(a),(b)], but smeared on a ring with radius $\tau = 2\pi/\ell$ \footnote{The rocking scans reveal a considerable broadening of the helical Bragg peaks on the surface of a sphere with radius $\tau=2\pi/\ell$}. A slight increase of the Fe concentration to $x$ = 0.10 considerably enhances the broadening, indicating an ill-defined orientation of the helix and a weakening of the magnetic anisotropy, which is possibly due to the increased chemical disorder \cite{grigoriev2009b}. 

The long-range helimagnetic order with a well-defined propagation direction disappears for $x$ $>$ $x^*$.  For $x$ = 0.11 and 0.14, broad isotropic rings of scattering instead of Bragg peaks are observed down to the lowest temperature measured, which are well below the respective transition temperatures inferred from susceptibility measurements [Table \ref{table}]. These rings of scattering intensify but remain broad with decreasing temperature, thus indicating finite helimagnetic correlations.

The effect of dilution on the zero-field helimagnetic order is further illustrated by Figure \ref{S_Q}(a), which displays the normalized scattering function $S(Q)$ at $\mu_0H$ = 0~T and below the transition temperature for several compositions. $S(Q)$ is obtained by radial averaging the SANS patterns of Fig. \ref{Zero_field_patterns} and is thus a one-dimensional representation of the two-dimensional scattering patterns. The lineshape of $S(Q)$ is dramatically different for $x$ $>$ $x^*$ than for $x$ $<$ $x^*$. Indeed, for $x$ $\leq$ 0.10, $S(Q)$ has a Gaussian lineshape with a constant Full Width Half Maximum (FWHM) of $\Delta Q$/$Q$ $\approx$ 0.16 that roughly corresponds to the resolution of the instrument. On the other hand, for $x$ = 0.11 and especially for $x$ = 0.14, $S(Q)$ is no longer resolution limited but broad, as indicated by the respective FWHM of $\Delta Q$/$Q$ $\approx$ 0.23 for $x$ = 0.11 and $\Delta Q$/$Q$ $\approx$ 0.65 for $x$ = 0.14. As further illustrated by Figs. \ref{S_Q}(b)-(d), which display $S(Q)$ at different temperatures for $x$ = 0.09, 0.11 and 0.14,  $S(Q)$ remains for $x$ $>$ $x^*$ broad down to the lowest temperature measured (1.8~K), indicating that helimagnetic correlations with finite correlation lengths persist to the lowest temperatures. 

We consider the Ornstein-Zernike relation to extract estimates for the correlation length $\xi$ and the pitch of the helix $\ell$:

\begin{equation}
S(Q)={C}/\left(\left(Q-2\pi/\ell\right)^2+\xi^{-2}\right),
\label{eq:OZ}
\end{equation}

\noindent with $C$ the Curie constant. The temperature dependence of both $\ell$ and $\xi$ are displayed in Fig. \ref{Fit_SQ}. For the sake of clarity, the values of $\ell$ at $T$ = 2.5~K and at $T_C$ are also provided in Table \ref{table}. 

Fig. \ref{Fit_SQ}(a) shows that $\ell$ decreases monotonously  with increasing Fe concentration: at $T$ = 2.5~K, it decreases from 18.2~nm for $x$ = 0 to 6.2~nm for $x$ = 0.14. As further addressed in the discussion, this indicates a strengthening of the DM interaction with respect to the ferromagnetic exchange interaction.

The correlation length follows a qualitatively similar behavior for all compositions with $x$ $<$ $x^*$: it increases monotonously with decreasing temperature and reaches approximately the pitch of the helix at $T_C$. On the other hand, for $x$ $>$ $x^*$, $\xi$ increases monotonously down to the lowest temperature. In particular for $x$ = 0.14, $\xi$ is significantly smaller than $\ell$ and reaching 5~nm $\sim$ 2/3$\ell$ at 1.8~K.

%############ Field Patterns ###########################################
\begin{figure*}[tb]
\begin{center}
\includegraphics[width= 1\textwidth]{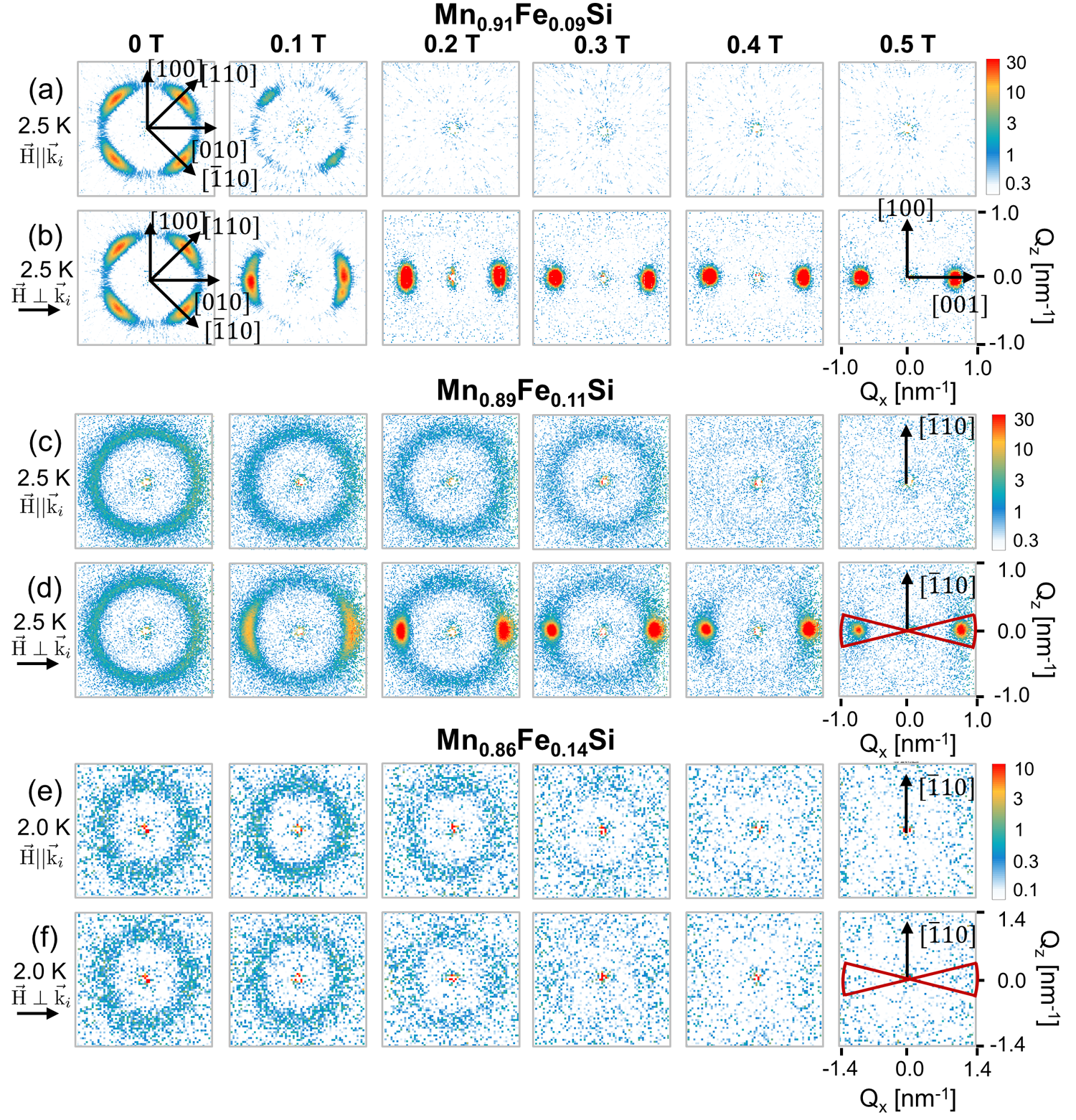}
\caption{SANS patterns measured at $T$ $<$ $T_C$ as a function of the magnetic field for Mn$_{1-x}$Fe$_x$Si with (a),(b) $x$ = 0.09 at $T$ = 2.5~K,  (c),(d) $x$ = 0.11 at $T$ = 2.5~K and (e),(f) $x$ = 0.14 at $T$ = 2.0~K. The magnetic field was applied (a),(c),(f) parallel to the incoming neutron beam ($\vec{H} || \vec{k}_i$) and (b),(d),(e) perpendicular to it ($\vec{H} \perp \vec{k}_i$). The two 30$^\circ$ red wedges in Panels (d) and (f) indicate the section of the detector over which the radial integration was performed to obtain $S(Q)$ for $\vec{H} \perp \vec{k}_i$. }
\label{Patterns_field}
\end{center}
\end{figure*}
%##########################################################################

\section{Magnetic Field}

\subsection{Measurements at $T$ = 2~K}

%############ S(Q) ###########################################
\begin{figure*}[tb]
\begin{center}
\includegraphics[width= 0.85\textwidth]{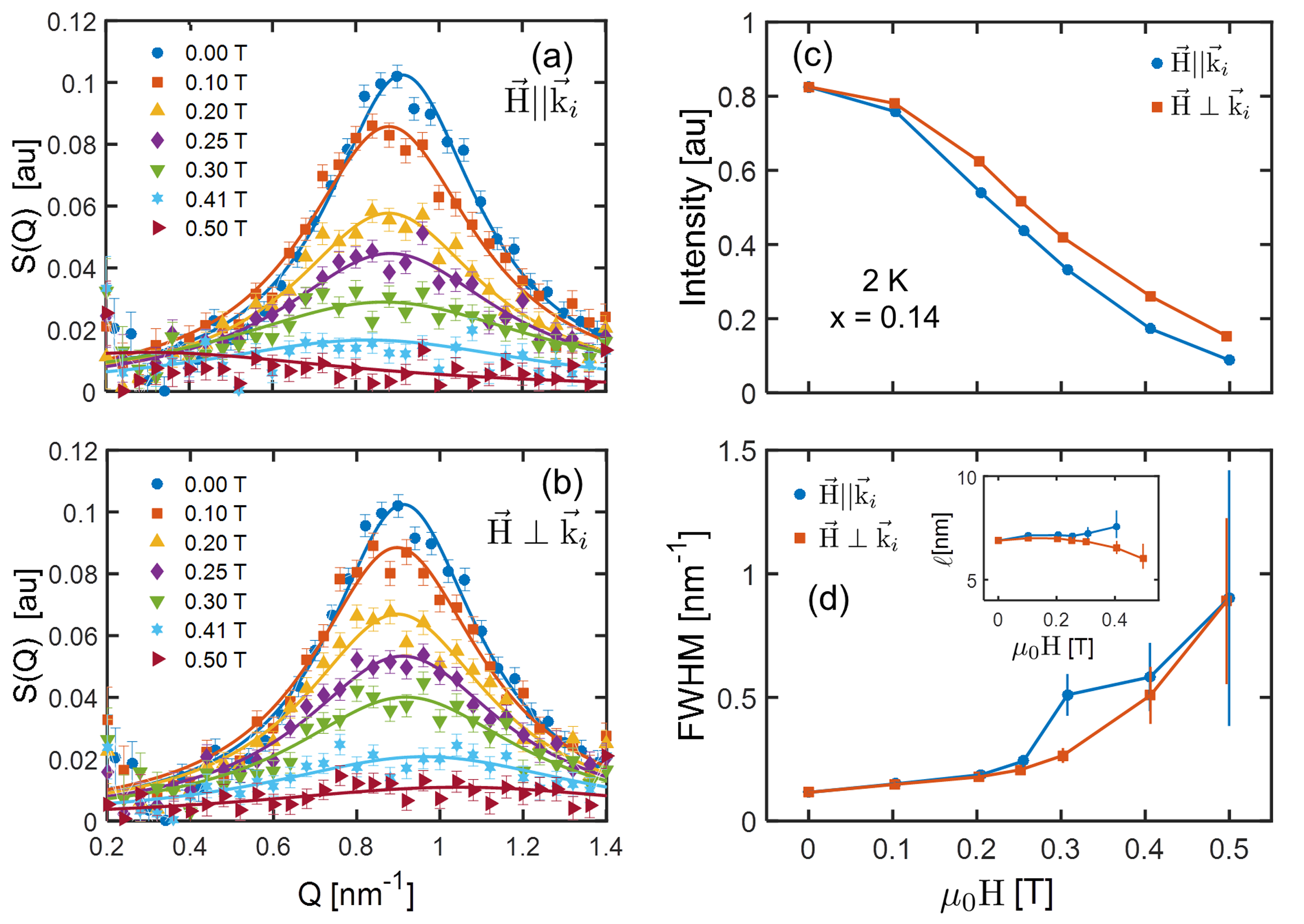}
\caption{SANS results under magnetic field for Mn$_{0.89}$Fe$_{0.14}$Si at $T$ = 2.0~K. Panel (a) shows the scattering function $S(Q)$ in arbitrary units obtained by radial averaging the scattered intensity over the entire detector with the magnetic field applied parallel to the incoming neutron beam ($\vec{H} || \vec{k}_i$). Panel (b) shows $S(Q)$ in arbitrary units with the magnetic field applied perpendicular to the incoming neutron beam ($\vec{H} \perp \vec{k}_i$). In the latter case, $S(Q)$ is deduced from radial averaging the scattered intensity over the two 30$^\circ$ wedges along the magnetic field direction indicated in Fig. \ref{Patterns_field}(f). Panel (c) shows the magnetic field dependence of the total scattered intensity obtained by summing the intensity over the entire detector for $\vec{H} || \vec{k}_i$ and over the two 30$^\circ$ wedges along the magnetic field for $\vec{H} \perp \vec{k}_i$. Panel (d) shows the magnetic field dependence of the Full-Width-Half-Maximum of $S(Q)$ and the inset shows the field dependence of the pitch of the helix $\ell$ as obtained from fitting the data of (a),(b) to eq. \ref{eq:OZ}. These fits are indicated by the solid lines in Panels (a) and (b).}
\label{0p14_field}
\end{center}
\end{figure*}
%##########################################################################

Figure \ref{Patterns_field} displays SANS patterns recorded as a function of magnetic field for Mn$_{1-x}$Fe$_x$Si with (a),(b) $x$ = 0.09, (c),(d) $x$ = 0.11 at $T$ = 2.5~K and (e),(f) $x$ = 0.14 at $T$ = 2~K, i.e. for $T$ $<$ $T_C$. These patterns have been collected in two complementary experimental configurations: one with the magnetic field parallel to the incoming neutron beam ($\vec{H} || \vec{k}_i$) and one with the field perpendicular to the incoming neutron beam ($\vec{H} \perp \vec{k}_i$).

Fig. \ref{Patterns_field}(a) shows SANS patterns for $x$ = 0.09 with $\vec{H} || \vec{k}_i$ i.e. the configuration that is sensitive to helical modulations perpendicular to the magnetic field. At zero magnetic field, the pattern displays four smeared Bragg peaks oriented along the  $\langle 110 \rangle$  crystallographic directions that originate from the helical phase. For $\mu_0H$ = 0.1~T two of these peaks disappear completely, while the other two considerable weaken in intensity. For magnetic fields exceeding 0.1~T, the scattered intensity is negligible, as expected for the conical phase where all helices are oriented along the magnetic field.

A complementary picture is provided by the SANS patterns of Fig. \ref{Patterns_field}(b) measured with $\vec{H} \perp \vec{k}_i$, an experimental configuration which probes helical modulations oriented along the magnetic field.  In this configuration, spots of scattered intensity are found along the horizontal field direction for $\mu_0H$ $\gtrsim$ 0.05~T. Together with the disappearance of the Bragg peaks for $\vec{H} || \vec{k}_i$, this behavior marks the conical phase in which the propagation direction of the helices changes from along the $\langle 110 \rangle$ crystallographic direction to the direction of the magnetic field. These Bragg peaks disappear for $\mu_0H$ $\gtrsim$ 0.6~T, marking the onset of the field polarized state (not shown).

The patterns for $x$ = 0.11, which are displayed in Fig. \ref{Patterns_field}(c),(d), are qualitatively different from the ones for $x$ $<$ $x^*$. They show at zero field the isotropic ring of scattering discussed in the previous section. This isotropic scattering gradually disappears when the magnetic field is applied along $\vec{k_i}$ [Fig. \ref{Patterns_field}(c)]. This is consistent with the patterns  for $\vec{H} \perp \vec{k}_i$ [Fig. \ref{Patterns_field}(d)], which show that the scattering concentrates along the field direction, ultimately leading to intense spots at high magnetic fields. However, the alignment of the helices along the magnetic field is very gradual and the intense Bragg-like spots coexist with a weaker ring of scattering up to $\mu_0H$ = 0.5~T. For $\mu_0H$ $\gtrsim$ 0.6~T, the scattered intensity vanishes (not shown), indicating the onset of the field polarized state.

The alignment of the helimagnetic correlations by the magnetic field for $x$ = 0.14 is considerably weaker than for $x$ = 0.11. Although a slightly larger fraction of the helimagnetic correlations is still oriented along the magnetic field for $\mu_0H$ $\geq$ 0.20~T, this effect is much smaller than for $x$ = 0.11. The effect of the magnetic field is also seen on the scattering function $S(Q)$, displayed in Figure \ref{0p14_field}(a) and (b) for $T$ = 2.0~K. $S(Q)$ is obtained by radial averaging over the entire detector for $\vec{H} || \vec{k}_i$ [Fig. \ref{0p14_field}(a)], and over the two wedges of $\pm$ 15$^\circ$ around the magnetic field direction, indicated in Fig. \ref{Patterns_field}(d), for $\vec{H} \perp \vec{k}_i$ [Fig. \ref{0p14_field}(b)]. The results show that for both experimental configurations, $S(Q)$ broadens considerably with increasing magnetic fields and decreases in intensity. 

The decrease of the scattered intensity of $S(Q)$ is also seen in the magnetic field dependence of the total scattered intensity displayed in Fig. \ref{0p14_field}(c), and which is more pronounced for $\vec{H} || \vec{k}_i$ than for $\vec{H} \perp \vec{k}_i$. This decrease can be the result of either the suppression of the helimagnetic correlations as a whole, and/or the orientation of the magnetic moments within the helices towards the magnetic field, as is also the case in the conical phase for $x$ $<$ $x^*$. In addition, $S(Q)$ broadens for $x$ = 0.14 considerably with increasing magnetic field, as highlighted by the magnetic field dependence of the FWHM of $S(Q)$ displayed in Fig. \ref{0p14_field}(d). This behavior is very different from that found for lower Fe dopings and shows that the characteristic correlation length decreases substantially with increasing magnetic field: from $\sim$ 5~nm at $\mu_0H$ = 0~T to $\xi$ $\sim$ 2~nm or approximately 1/4 $\ell$ at $\mu_0H$ = 0.5~T. This indicates a complex magnetization process in which the magnetic field first breaks the longer helical correlations, which are possibly those that encompass the lowest degree of disorder. On the other hand, the shorter helices appear to be more robust, as is is also the case for vortices in re-entrant spin glasses \cite{mirebeau2018}. 

\subsection{Skyrmion Lattice Phase}
%############ SkL ###########################################################
\begin{figure*}[tb]
\begin{center}
\includegraphics[width= 0.8\textwidth]{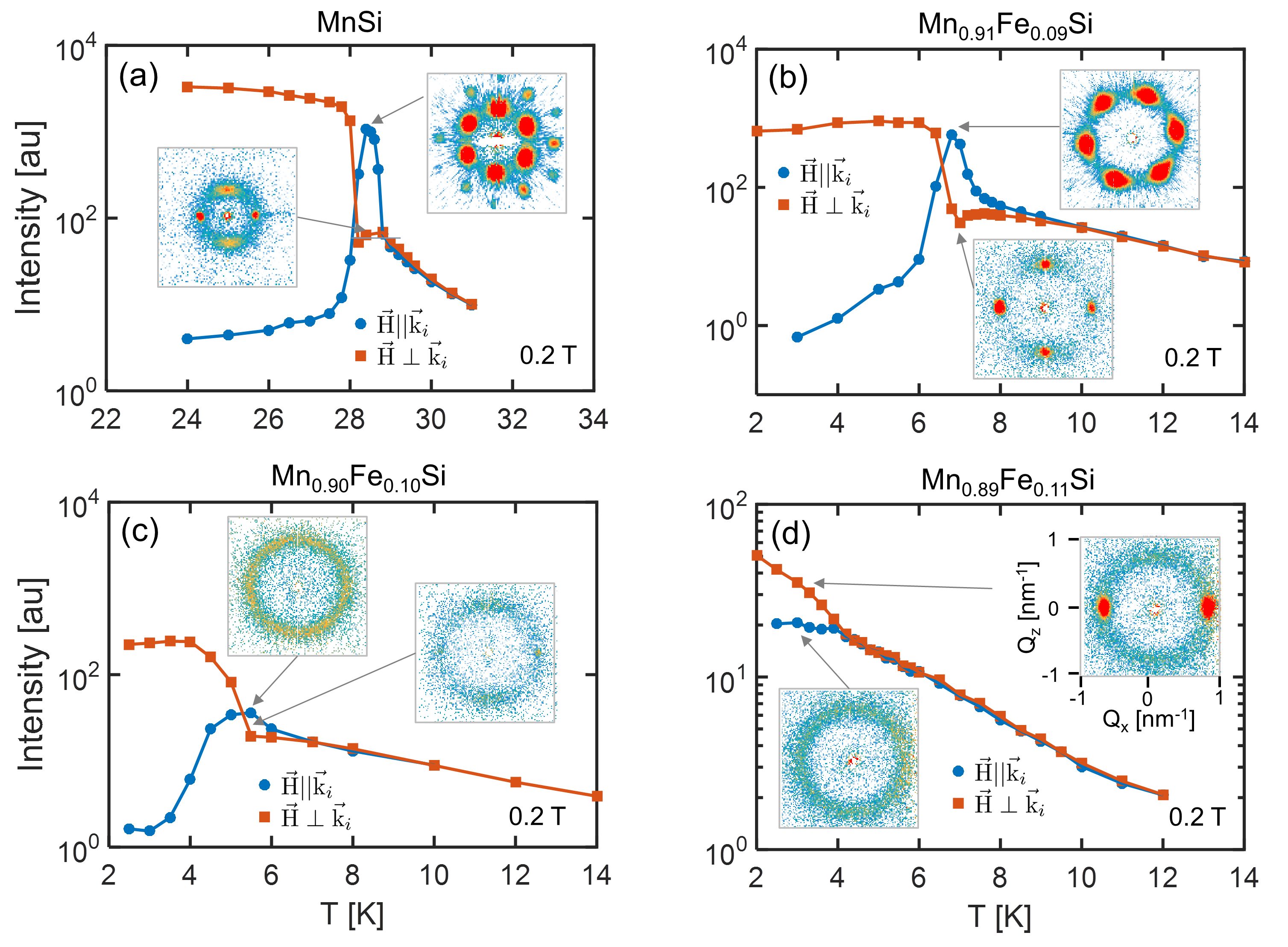}
\caption{Temperature dependence of the total scattered intensity at $\mu_0H$ = 0.20~T for Mn$_{1-x}$Fe$_x$Si with (a) $x$ = 0, (b) $x$ = 0.09, (c) $x$ = 0.10, and (d) $x$ = 0.11. The field was applied both parallel ($\vec{H} || \vec{k}_i$) and perpendicular ($\vec{H} \perp \vec{k}_i$) to the incoming neutron beam. The insets show characteristics SANS patterns at the indicated temperatures for both field configurations. }
\label{SkL}
\end{center}
\end{figure*}
%##########################################################################

The impact of Fe substitution on the skyrmion lattice correlations in Mn$_{1-x}$Fe$_x$Si is illustrated by Fig. \ref{SkL} which displays the total scattered intensity for four different Fe concentrations as a function of temperature at $\mu_0H$ = 0.20~T, i.e. in the heart of the skyrmion lattice phase. The magnetic field was applied both parallel ($\vec{H} || \vec{k}_i$) and perpendicular ($\vec{H} \perp \vec{k}_i$) to the incoming neutron beam. 

For all compositions, the intensity above $T_C$ is almost the same for both experimental configurations, as expected for isotropic correlations. In contrast, at low temperatures the intensity for $\vec{H} \perp \vec{k}_i$ is significantly higher than the one for $\vec{H} || \vec{k}_i$, as expected for the conical phase. In the intermediate temperature region, the onset of skyrmion lattice correlations leads to a different and non-monotonic evolution of the scattered intensity with temperature.

MnSi and $x$ = 0.09 show basically the same behavior with a sharp maximum in intensity for $\vec{H} || \vec{k}_i$ occurring in a temperature region of about $\pm$1~K below $T_C$. Together with the characteristic six-fold symmetry of the scattering displayed in the insets of Figs. \ref{SkL}(a) and \ref{SkL}(b), this maximum marks the skyrmion lattice phase \cite{muhlbauer2009,seki2015skyrmions,bauer2016generic}. We note that skyrmion lattice correlations coexist with conical correlations, as seen from the finite amount of scattering in the configuration where $\vec{H} \perp \vec{k}_i$.

With dilution, the skyrmion lattice scattering weakens considerably. In addition, for $x$ = 0.10 an isotropic ring of scattering instead of six clear peaks is found. This is similar to Fe$_{1-x}$Co$_{x}$Si \cite{munzer2010,bannenberg2016}, and the absence of clear peaks is likely related to a weakening of the fourth and sixth order cubic anisotropy terms responsible for the alignment of the skyrmion lattice with respect to the crystallographic one \cite{muhlbauer2009,munzer2010,bannenberg2017reorientations}. The disappearance of Bragg peaks in the skyrmion lattice phase is also consistent with the broadening of the helical Bragg peaks [Fig \ref{Zero_field_patterns}], which also indicates a weakening of the anisotropy with increasing Fe substitution. 

A substantially different behavior unfolds for $x$ = 0.11, where there are no clear indications for skyrmion lattice correlations [Fig. \ref{SkL}(d)]. These results are consistent with magnetic susceptibility measurements \cite{bannenberg2018mnfesisquid} which for $x$ $>$ $x^*$ do not show any indication of a skyrmion lattice phase. On the other hand, the sizable topological Hall effect reported for this composition \cite{franz2014} possibly indicates the existence of individual skyrmions or clusters of skyrmions in this region of the magnetic phase diagram.

%############ Contour plot par ###########################################
\begin{figure*}[tb]
\begin{center}
\includegraphics[width= 0.8\textwidth]{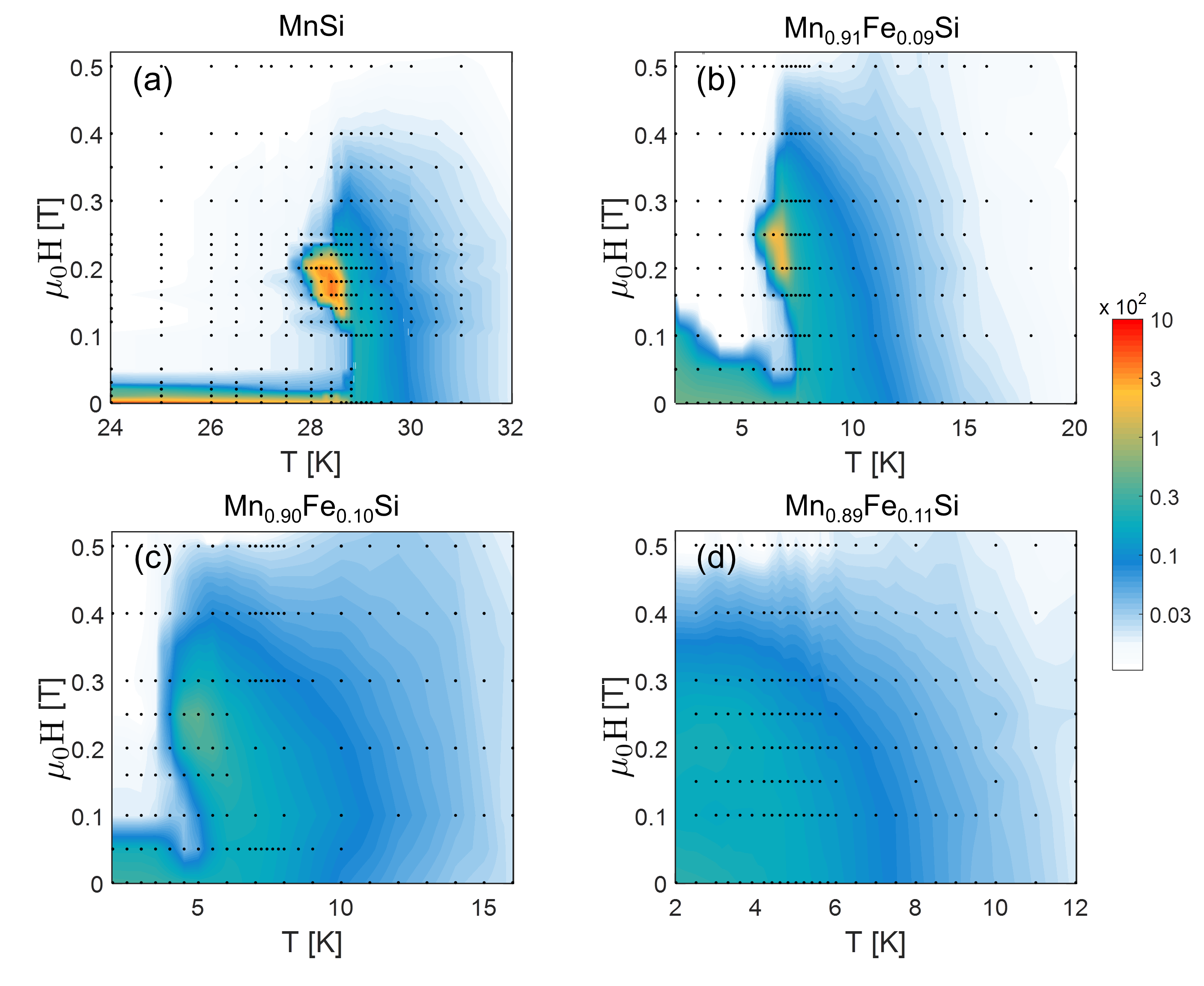}
\caption{Contour plots showing the total scattered SANS intensity in arbitrary units in the configuration where the magnetic field was applied parallel to the incoming neutron beam ($\vec{H} || \vec{k}_i$) for Mn$_{1-x}$Fe$_x$Si with (a) $x$ = 0, (b) $x$ = 0.09, (c) $x$ = 0.10, and (d) $x$ = 0.11. The black dots indicate the points at which the SANS measurements were performed.}
\label{Contour_par}
\end{center}
\end{figure*}
%##########################################################################

%############ Contour plot perp ###########################################
\begin{figure*}[tb]
\begin{center}
\includegraphics[width= 0.8\textwidth]{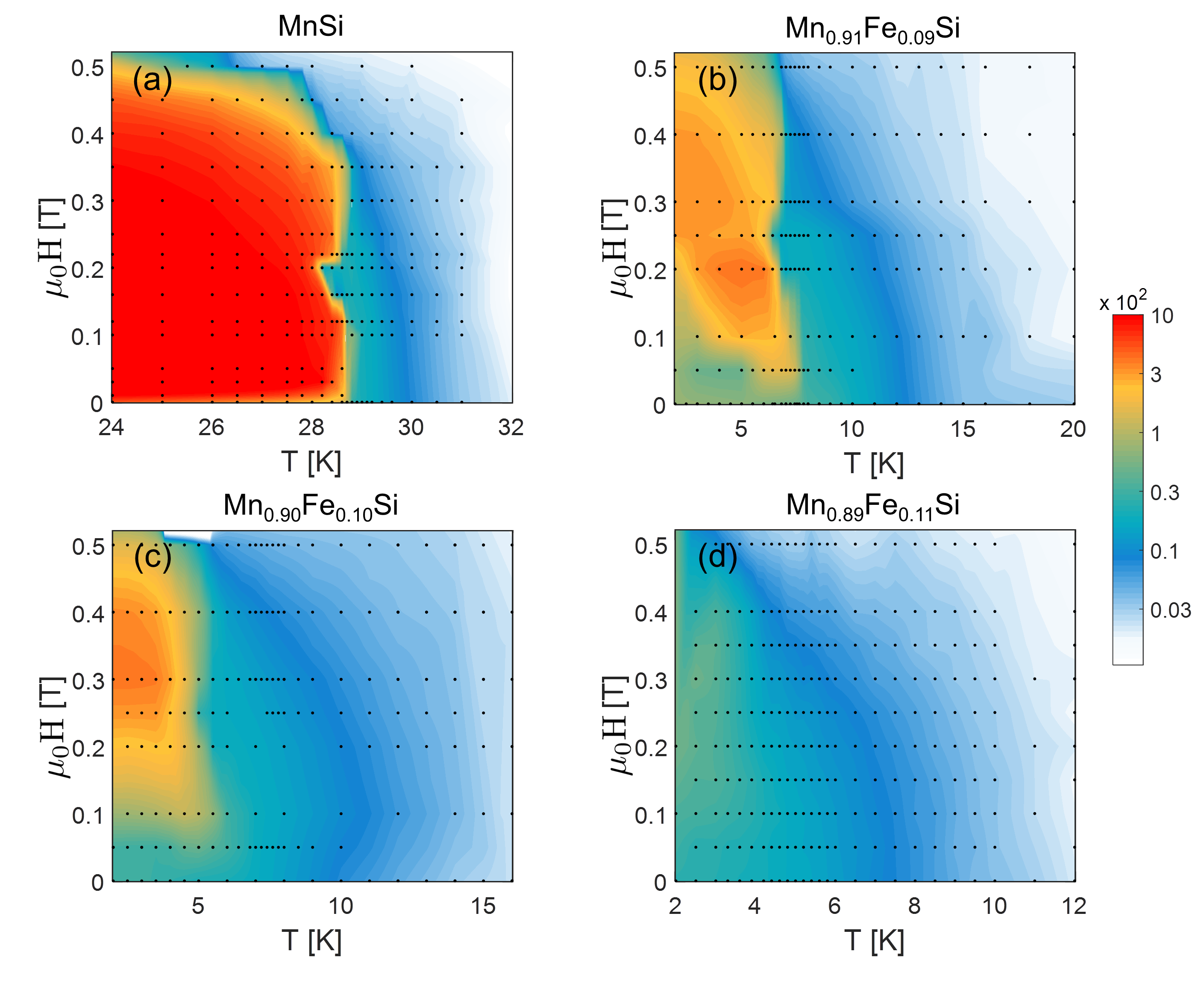}
\caption{Contour plots showing the total scattered SANS intensity in arbitrary units  in the configuration where the magnetic field was applied perpendicular to the incoming neutron beam ($\vec{H} \perp \vec{k}_i$) for Mn$_{1-x}$Fe$_x$Si with (a) $x$ = 0, (b) $x$ = 0.09, (c) $x$ = 0.10, and (d) $x$ = 0.11. The black dots indicate the points at which the SANS measurements were performed.}
\label{Contour_perp}
\end{center}
\end{figure*}
%##########################################################################

%############ Doping ###########################################
\begin{figure}[tb]
\begin{center}
\includegraphics[width= 0.45\textwidth]{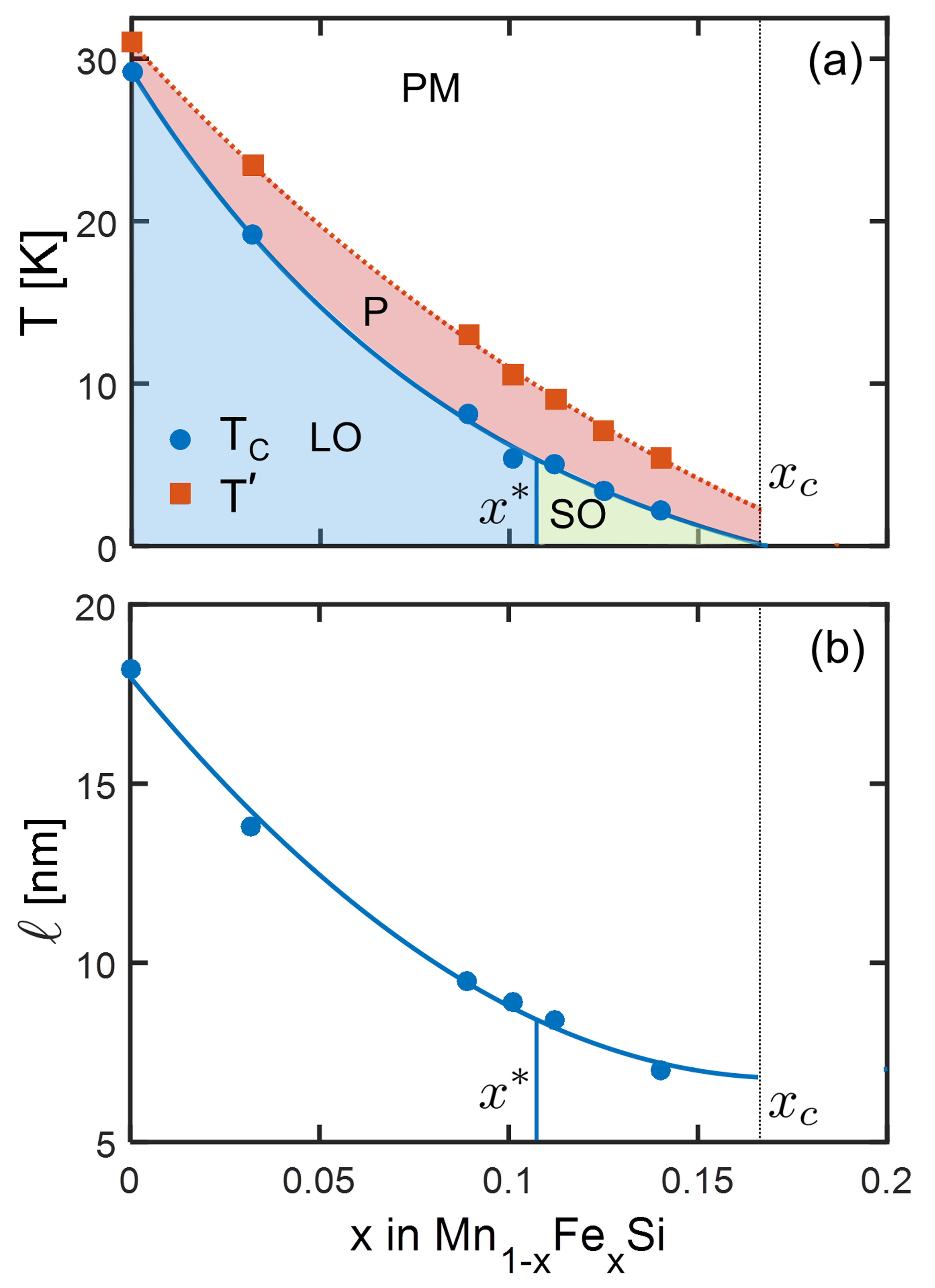}
\caption{Fe concentration dependence of (a) the critical temperature $T_C$ and $T^\prime$ and (b) the pitch of the helix $\ell$. The critical temperature $T_C$ and $T^\prime$, which marks the onset of the (short-ranged) helimagnetic correlations in the precursor phase are adapted from ref. \cite{bannenberg2018mnfesisquid}. LO indicates long-range helimagnetic correlations, SO short-range helimagnetic correlations, P the precursor phase and PM the paramagnetic phase.}
\label{Doping}
\end{center}
\end{figure}
%##########################################################################

\subsection{Phase Diagrams}
An overview of the effect of the Fe concentration on the magnetic field - temperature phase diagram of several Mn$_{1-x}$Fe$_x$Si compositions is provided by Figures \ref{Contour_par} and \ref{Contour_perp}, which display contour plots of the scattered intensity as a function of temperature and magnetic field for both $\vec{H} || \vec{k}_i$ and $\vec{H} \perp \vec{k}_i$. The contour plots are in excellent agreement with the phase diagrams published in ref. \cite{bannenberg2018mnfesisquid} which are derived from magnetic susceptibility measurements.

Fig. \ref{Contour_par}(a) shows the contour plot of the scattered intensity of MnSi with $\vec{H} || \vec{k}_i$. At low magnetic fields and below $T_C$ $\sim$ 28.7~K, the high intensity originates from the helical phase. Magnetic fields beyond 0.05~T align the helices towards their direction, leading to almost zero scattered intensity in this configuration. On the other hand, the onset of the conical phase is accompanied by a strong increase in intensity in the complementary experimental configuration with $\vec{H} \perp \vec{k}_i$ [Fig. \ref{Contour_perp}(a)]. In this configuration, the scattered intensity of the conical phase is suppressed for $\mu_0H$ $\gtrsim$ 0.5~T, i.e. in the field polarized state in which the magnetic field aligns the magnetic moments along its direction.

Just below $T_C$ and in a magnetic field range of 0.14 - 0.22~T, a pocket of increased intensity shows up in the contour plot for $\vec{H} || \vec{k}_i$. As discussed in the previous section, the increased intensity in this pocket, also known as the A-phase, originates from skyrmion lattice correlations that orient perpendicular to the applied magnetic field. In the complementary configuration of Fig. \ref{Contour_perp}(a), a clear decrease in intensity is seen in this region of the phase diagram, indicating a partial suppression of the conical correlations. 

The magnetic field - temperature contour plots of the scattered intensity for $x$ = 0.09 and $x$ = 0.10 look qualitatively similar to each other and to the ones for MnSi. However, some subtle differences are visible. Similar to Fe$_{1-x}$Co$_x$Si \cite{munzer2010,bauer2016,bannenberg2016,bannenberg2016squid}, and consistent with susceptibility measurements \cite{bauer2010,bannenberg2018mnfesisquid}, the transition line from the helical to the conical phase is no longer horizontal, but shifts to higher magnetic fields with decreasing temperature. Furthermore, the increase of the scattered intensity for $\vec{H} || \vec{k}_i$ in the A-phase is less pronounced, representing a weakening of the skyrmion lattice correlations. This weakening is accompanied by a relative increase of the intensity originating from conical correlations that coexist with the skyrmion lattice correlations. 

The contour plots for $x$ = 0.11 bear some similarities to the ones for $x$ $<$ $x^*$ but also reveal substantial differences. The scattered intensity for $\vec{H} || \vec{k}_i$ [Fig. \ref{Contour_par}(d)] persists until the lowest temperature measured and for magnetic fields up to 0.45~T. In the complementary configuration of Fig. \ref{Contour_perp}(d) with $\vec{H} \perp \vec{k}_i$, the intensity is enhanced by the magnetic field for temperatures lower than $T_C$ $\lesssim$ 5~K. These results indicate that the helimagnetic correlations are only partly oriented along the magnetic field, even at relatively large fields. In addition, the contour plots provide no indication of a skyrmion lattice phase.

Above $T_C$, all contour plots show a similar gradual decrease of the scattered intensity with increasing temperature and for all magnetic fields up to $\mu_0H$ $\sim$ 0.3~T. For higher fields, this intensity decreases gradually in the configuration with $\vec{H} || \vec{k}_i$, whereas it persists for the complementary one. This is not surprising, as sufficiently large magnetic fields suppress or align the helimagnetic correlations along their direction \cite{pappas2017}. The intensive scattering above $T_C$ defines the precursor region, which occurs over a temperature range that increases substantially with increasing Fe doping for $x$ $<$ $x^*$ \cite{grigoriev2009b,bauer2010,grigoriev2011,bannenberg2018mnfesisquid}.

%############################################################################
\section{Discussion}
%############################################################################
The experimental results are summarized in the concentration dependence of the transition temperature, derived from previous susceptibility measurements \cite{bannenberg2018mnfesisquid}, and the pitch of the helix $\ell$ depicted in Fig. \ref{Doping}. These plots show that the transition temperature vanishes at $x_C$ $\approx$ 0.17 and that Fe substitution leads to a significant reduction of the helimagnetic pitch $\ell$. In addition, a transition from long-ranged to short-ranged helimagnetic correlations occurs already at $x^*$ $\approx$ 0.11.

In order to understand the evolution of the helimagnetic order with $x$, we consider the free energy per unit cell of a cubic chiral magnet such as Mn$_{1-x}$Fe$_x$Si. This free energy is the sum of the ferromagnetic interaction, DM  interaction, Zeeman and magnetic anisotropy energies and is given by:

\begin{equation}
\begin{aligned}
f = \frac{Ja^2}{2}\sum_{i=x,y,x}{\left[\partial_i\hat{m}\cdot\partial_i\hat{m}\right]}+&aD\hat{m}\cdot\nabla\times\hat{m}\\
&-a^3\mu_0M\hat{m}\cdot\vec{H}+f_a,
\end{aligned}
\end{equation}

\noindent with $a$ the lattice constant, $\hat{m}$ the unit vector in the direction of the magnetization, $M$ the magnetization value, and $J$ and $D$ the strength of the ferromagnetic exchange and DM interaction, respectively, and $f_a$ the magnetic anisotropy contribution \cite{qian2018,bauer2016generic,seki2015skyrmions}. In this model, the transition temperature is proportional to $J$. By substituting the conical spiral Ansatz, i.e. $\hat{m}$ = $\cos \theta \hat{e_3}+\sin \left[ \theta \cos(\vec{\tau} \cdot \vec{x})\hat{e_1}+\sin(\vec{\tau} \cdot \vec{x})\hat{e_2}\right]$, with $\theta$ the cone angle, $\vec{\tau}$ the helical propagation vector and ($\hat{e_1},\hat{e_2},\hat{e_3}$) a set of orthogonal unit vectors, one can derive that the pitch of the helical modulation is, in the absence of Zeeman and magnetic anisotropy, given by: $\ell$ = $\frac{2\pi}{a}\frac{J}{D}$. Moreover, the field at which the conical-to-field polarized transition occurs at $T$ = 0~K is $\mu_0H_{C2}(0 \text{K})$ = $\frac{D^2}{Ja^3M}$ \cite{qian2018}. 

By comparing these expressions with the experimental results, we obtain that $J$ $\rightarrow$ 0 for $x$ $\rightarrow$ $x_c$. On the other hand, $\ell$ is reduced from $\sim$ 18~nm for $x$ = 0.0 to $\sim$ 6 nm at $x$ = 0.14, i.e. by a factor of $\approx$ 3, implying that $D/J$ and thus the relative strength of the DM interaction increases with dilution. The DM interaction itself originates from the anisotropic exchange between the spins of the magnetic atoms and is thus a first-order correction to the Heisenberg exchange in spin-orbit coupling $\lambda$: $D \propto J\frac{\lambda}{\Delta}$, with $\Delta$ the typical electron excitation energy on the site of the magnetic atom \cite{M,qian2018}.  Thus, $D/J \propto \frac{\lambda}{\Delta} \propto \ell$, and although $J$ vanishes at $x_C$, the ratio of $D/J$ should remain finite, which is in good agreement with the experimental findings. Similar considerations explain why $\mu_0H_{C2}(0 \text{K})$ does not increase substantially as $x$ $\rightarrow$ $x_C$.

The helimagnetic order is already affected at dopings well below  $x_c$. The helimagnetic spiral reorients at $x$ = 0.09 from the $\langle 111 \rangle$ to the $\langle 110 \rangle$ crystallographic directions. Such a reorientation cannot be explained from the anisotropy term responsible for $\vec{\tau}$ $||$ $\langle 111 \rangle$ at $x$ = 0, $f_{a1} = K(m_x^4+m_y^4+m_z^4)$, as this term only has minima for $\vec{\tau}$ $||$ $\langle 111 \rangle$ for $K$ $<$ 0 and $\vec{\tau}$ $||$ $\langle 100 \rangle$ for $K$ $>$ 0 \cite{bak1980}. It therefore implies that other anisotropic terms become relevant. In fact, such a deviation from $\vec{\tau}$ $||$ $\langle 100 \rangle$ or $\langle 111 \rangle$ is not unique to Mn$_{1-x}$Fe$_x$Si but has been reported for the partially ordered state in MnSi under hydrostatic pressure \cite{pfleiderer2004,kruger2012}. 

As the helices do not align along a specific crystallographic lattice direction for $x$ $>$ $x^*$, we deduce that either the magnetic anisotropy is very weak or that disorder smears the effect of anisotropy. However, solely the differences in the anisotropy cannot explain the quantitatively different behavior for $x$ $>$ $x^*$. Indeed, the correlation length remains finite for these dopings down to the lowest temperature and magnetic fields do not raise the directional degeneracy as they do not align all helices along their direction. Differences between the magnetic behavior have also been reported based on measurements of bulk quantities \cite{demishev2013,demishev2014,bannenberg2018mnfesisquid} and it has been suggested that the crossover at $x^*$ is due to quantum fluctuations that destabilize the long-range helimagnetic order \cite{demishev2013,glushkov2015,demishev2016a,demishev2016b}. In this approach, $x^*$ would be a quantum critical point.  We conjecture that this is not necessarily  the case as the disappearance of long-range helimagnetic correlations might be due to chemical disorder. This hypothesis is supported by the data obtained under magnetic field [Fig. \ref{0p14_field}] that show a complex magnetization process. In this process, the magnetic field first breaks the longer helices, possibly those that encompass the lowest degree of disorder, whereas the shorter helices are more robust.

The isotropic zero magnetic field SANS patterns in combination with the finite helical correlations for $x$ $>$ $x^*$ $\approx$ 0.11 indicate that $x^*$ is associated with the disappearance of long-range periodic helimagnetic order and that the short-ranged helimagnetic correlations for $x$ $>$ $x^*$ are completely degenerate in space. However, from these results it is not clear whether this helimagnetic state fluctuates, as in the precursor phase in MnSi above $T_C$ \cite{pappas2009,pappas2017}, or not.  The final answer can be given by additional inelastic neutron scattering or muon spin spectroscopy experiments.

\section{Conclusion}
In conclusion, the results show that the helimagnetic order in Mn$_{1-x}$Fe$_x$Si is suppressed with increasing Fe content. The long-range helimagnetic correlations, which reorient at zero magnetic from $\langle 111 \rangle$ at low Fe concentrations towards $\langle 110 \rangle$ at $x$ = 0.09, disappear completely for $x$ $>$ $x^*$. The helices have for $x$ $>$ $x^*$ finite lengths, are completely degenerate in space, and bear similarities to those found in the precursor phase of MnSi. Magnetic fields gradually suppress and partly align these helices along their direction trough a complex magnetization process. 

\begin{acknowledgments}
The authors wish to thank the ISIS support staff for their assistance and are grateful for the kind help of G. Stenning and R. Perry with aligning the single crystals and B. Pedersen for testing our initial crystals with Neutron Laue Diffraction at FRMII. R.W.A. Hendrikx at the Department of Materials Science and Engineering of the Delft University of Technology is acknowledged for the XRF analysis. M. Mostovoy is acknowledged for fruitful discussions.  D. Bosma is acknowledged for EDS/SEM analysis of  the single crystals. Experiments at the ISIS Pulsed Neutron and Muon Source were supported by a beamtime allocation from the Science and Technology Facilities Council and The Netherlands Organization for Scientific Research (NWO). The work of LB and CP is financially supported by The Netherlands Organization for Scientific Research through project 721.012.102 (Larmor). FW was supported by the Helmholtz Society under contract VH-NG-840.
\end{acknowledgments}

%############################################################################
%\section{References}
%############################################################################
\bibliography{MnSi_Cu2OSeO3_Rotation}


\end{document}